\documentclass[]{aastex63}

\usepackage{natbib}
\usepackage{graphicx}
\usepackage{amsmath}
\usepackage{multirow}

\revised{\today}
\submitjournal{PSJ}

\shorttitle{GDR2 Masses}
\shortauthors{Murray et al.}
\graphicspath{{./}{figures/}}

\begin{document}

\title{The Challenge of Measuring Asteroid Masses with Gaia DR2 astrometry}

\correspondingauthor{Zach~Murray}
\email{zachary.murray@cfa.harvard.edu}
\author[0000-0002-8076-3854]{Zachary Murray}
\affil{Center for Astrophysics | Harvard \& Smithsonian, 60 Garden Street, Cambridge, MA 02138}

\begin{abstract}
    The Gaia second data release contains high-accuracy astrometric measurements of thousands of solar system bodies.  These measurements raise the possibility of determining asteroid masses by modeling scattering events between massive objects observed by Gaia. In this paper, we identify promising encounters between small asteroids that occur during the second data release and quantify the various errors involved in mass determination. We argue that in the best case, Gaia astrometry can provide constraints as tight as $\approx 1~\mathrm{km}$ on the positions of asteroids.
    Further, we find that even with general relativistic corrections, integrations of the solar system accumulate 1 km errors after 700 days. While not a problem for modeling DR2 astrometry, future Gaia data releases may require models accounting for additional effects such as gravitational harmonics of the sun and planets.  Additionally, due to sub-optimal astrometric uncertainty, the geometry of the observations, and the Gaia observing pattern result in much looser constraints in most cases, with constraints being several orders of magnitude weaker in some cases.  This suggests that accurate mass determination for the smallest asteroids will require additional observations, either from future Gaia data releases or from other sources.  We provide a list of encounters that are most promising for further investigation.
\end{abstract}


\section{Introduction}

There are several ways to determine the mass of an asteroid.  The most direct method is by measuring the gravitational effect of an asteroid on a nearby spacecraft.  This method has been used for nearly 30 years, examples include the NEAR flyby of 253 Mathilde \citep{Yeomans_1997}, the Hayabusa flyby of 25143 Itokawa \citep{Fujiwara_2006}, the Rosetta flyby of  21 Lutetia \citep{Patzold_2011}, and Dawn spacecraft's flybys of 4 Vesta \citep{Russell_2012} and 1 Ceres \citep{Park_2016}. Mass determinations from spacecraft are often highly accurate but are inherently limited due to the relatively small number of spacecraft that visit asteroids. 

There are also situations where the total mass of an asteroid pair can be determined from their orbital period and separation using Kepler's third law.  In the case that one of the two bodies is much smaller, the mass of the larger body can be accurately estimated.  This method has been has been successfully applied to near earth objects (NEOs) \citep{Ostro_2006,Shepard_2006}, main belt asteroids (MBAs) \citep{Marchis_2005}, centaurs \citep{Stansberry_2012} and trans-Neptunian objects (TNOs) \citep{Grundy_2009}. As noted in \citet{Carry_2012} this method of asteroid mass determination has been historically productive and has produced many highly accurate mass estimates.  The major weakness of this method, however, is that it requires a known and observable satellite object and high-resolution observations of the pair of objects.  This limits the estimates to a small fraction of asteroids. 

Another method used to estimate the masses of larger asteroids is to fit them to planetary ephemerides.   Hundreds of asteroids must be included to adequately model the available data~(see \citealt{Park_2021} for a recent example).  These mass measurements are less accurate than those achieved with spacecraft flybys, as well as what can typically be estimated with asteroid-asteroid encounters~\citep{Baer_2008}.  

Finally, the most productive method of mass determination is by measuring the gravitational influence of the massive asteroids on other small bodies.  This method has a long history, with one of the earliest measurements being done by \citet{Hertz_1966} of the asteroid Vesta.  These orbital deflection events typically occur between a larger primary asteroid and a smaller secondary asteroid, whose mass is effectively negligible.  The smaller asteroid is more affected by the encounter.  Detailed astrometric measurements of the scattered body can be used to infer the properties of the primary object \citep{Hilton_2002}.  However, these mass determinations often have large errors since the astrometric measurements of both bodies must be extremely accurate to properly model the interaction.  In addition, the influence of any other nearby bodies must be well known. 

The Gaia spacecraft was launched in 2013 and was primarily designed to make extremely accurate astrometric observations of Milky Way stars~\citep{GaiaCollaboration_2016}.  Despite it not being the spacecraft's primary mission, Gaia has also made observations of thousands of solar system objects, many of which also have extremely high astrometric precision.  In addition, the use of Gaia data to further constrain historical interactions has been demonstrated \citep{Siltala_2022}.  However, Gaia's extreme astrometric precision also presents the opportunity of measuring the masses of much smaller asteroids, whose encounters occurred recently and were observed by Gaia. Exploring this potential with Gaia Data Release 2 (DR2) is the subject of this paper. Recent advances in removing systematic errors in ground-based astrometry and correcting star-catalog biases may help further enlarge the collection of observational data that can be considered for extremely high precision work like that presented in this paper ~\citep{Farnocchia_2015,Fortino_2021}.

\section{The Physics of Gravitational Scattering}

The basic physics of gravitational scattering is well understood. For example, \citet{Hilton_2002} presents a first-order ballistic approximation of asteroid-asteroid encounters, showing that only the sum of the two asteroid masses can be computed. The deviation angle of a small asteroid being scattered by a larger one can be written as 
\begin{equation}
    \tan\left(\frac{\theta}{2} \right) = \frac{G(m+M)}{v^2 b},
    \label{eq:1}
\end{equation}
where $\theta$ is the deviation angle, $M$ the mass of the primary body, $m$ the mass of the scattered body, $b$ the impact parameter, and $v$ the relative velocity between the two bodies. Hence, the most useful encounters -- those which produce the largest observable deviations - are those with low relative velocities and small impact parameters. Most asteroids have small inclinations and hence a planar treatment is often appropriate.  For co-planar, relatively rapid encounters and to first order in eccentricity, encounters will change the semi-major axis or eccentricity of the scattered body~\citep{Danby_1988}.  Hence, as argued in \citet{Hilton_2002}, the main observable is, therefore, the total change in true longitude which increases roughly linearly in time. The longer one waits after a close encounter, the more sensitively one can constrain the dynamics of that encounter and, thus, the mass.  

Gaia achieves its extraordinary precision by accurately timing the transits of these stars across the detectors. The Gaia spacecraft rotates at a constant rate causing objects in its field of view to constantly drift along a given direction, the along-scan or ``AL'' direction). This results in the AL direction having extremely well-measured astrometry and the ``AC'' (across scan) direction being limited by the Gaia point spread function. This results in a peculiar error profile, with extremely accurate astrometry as good as $\approx 1 \textrm{mas}$ precision in the AL direction and  much looser constraints in the AC direction, as large as $100 \textrm{mas}$ precision, except for bright objects in which the precision is similar along both directions~\citep{GaiaCollaboration_2018}.  

When projected into equatorial coordinates the asymmetry in precision results in highly elongated  error ellipses. 

Despite this fantastic astrometric accuracy the Gaia observations do have a few limitations. First, the Gaia DR2 observations only span from JD 2456874 to JD 2457531. This represents a total time scale of just over 657 Julian Days - less than an orbital period for most main belt asteroids.  These time scales are much shorter than is typically used in mass determinations via scattering.  This has a major disadvantage as it does not allow for much time for changes in semi-major axis at the time of the encounter to accumulate into large observable angular displacements.  Hence, instead of looking for deviations of order an arcsecond over decades as explored in \citet{Hilton_1996} we instead look for scattering of order a milliarcsecond (mas) over just 2 years.   In addition to the short time span, the positions of solar system objects in Gaia DR2 must be corrected for relativistic light deflection due to the gravity of massive objects in the solar system.  This correction was done with the same procedure as used for Gaia stars and hence does not take into account the finite distance of solar system objects from the spacecraft.  This results in errors as large as $2~\mathrm{mas}$ on the Gaia astrometry \citep{GaiaCollaboration_2018}.   

Limitations aside, the Gaia DR2 dataset is vast and includes a total of 14099 Asteroids with around 2 million individual astrometric measurements across those bodies.  The vast majority of these bodies are main belt asteroids with a few Jupiter Trojans, NEOs, and TNOs. Solely by virtue of its size, such a data set will have many close encounters between individual asteroids although only a small fraction of those encounters will be strong enough to lead to deflections that are large enough that a mass can be inferred. Hence, searching for scattering events in the DR2 dataset requires that efficient filters be developed for separating promising encounter from the rest of the data set. 

\section{Filtering Interactions}    

We begin our filtering by conducting numerical integrations to determine which asteroids undergo close encounters, which we define as coming within $0.01 \mathrm{AU}$ of each other.  This criterion is adopted from \citet{Hilton_1996} and is ultimately arbitrary.  However, there are good reasons to believe it is wider than necessary in our case, as \citet{Hilton_1996} examined scattering by only the most massive asteroids.  Most of the bodies in our sample are significantly smaller. In our case overestimating in this way is desirable, as casting a wide net ensures we do not miss any potential encounters. 

We adopt a model consisting of the Sun, the eight major planets, and the 343 main belt asteroids that were used in the computation of the JPL DE430 and DE431 ephemerides \citep{Folkner_2014}.  We represent the Earth-Moon system with a particle at its barycenter with the total mass of the system.  We also exclude the Pluto-Charon system.  As our goal is to identify candidate encounters, small errors in ephemerides are acceptable.  Since high precision is unnecessary to construct an initial set of candidates we treat all the Gaia observed bodies not already accounted for in JPL as test particles and integrate the resulting system under strictly Newtonian dynamics with the \texttt{ias15} integrator adaptive integrator \citep{Everhart_1985,Rein_2015} implemented in \texttt{REBOUND} \citep{REBOUND}.  This results in a total of 1650 potentially interesting encounters - all between main belt asteroids.  

%try starting in center

While most of the encounters are of short duration, defined as the time elapsed where the bodies are within $0.01 \mathrm{AU}$ of each other, a few are much longer and can be as long as $50~\mathrm{days}$ in certain cases. As expected, we find many more distant encounters than very close ones. While most of these encounters are short, ballistic encounters, the longer encounters disproportionately constrain the masses of the bodies due to their larger deflections.  

\begin{figure}[!ht]
\centering
    \includegraphics[totalheight=7.5cm]{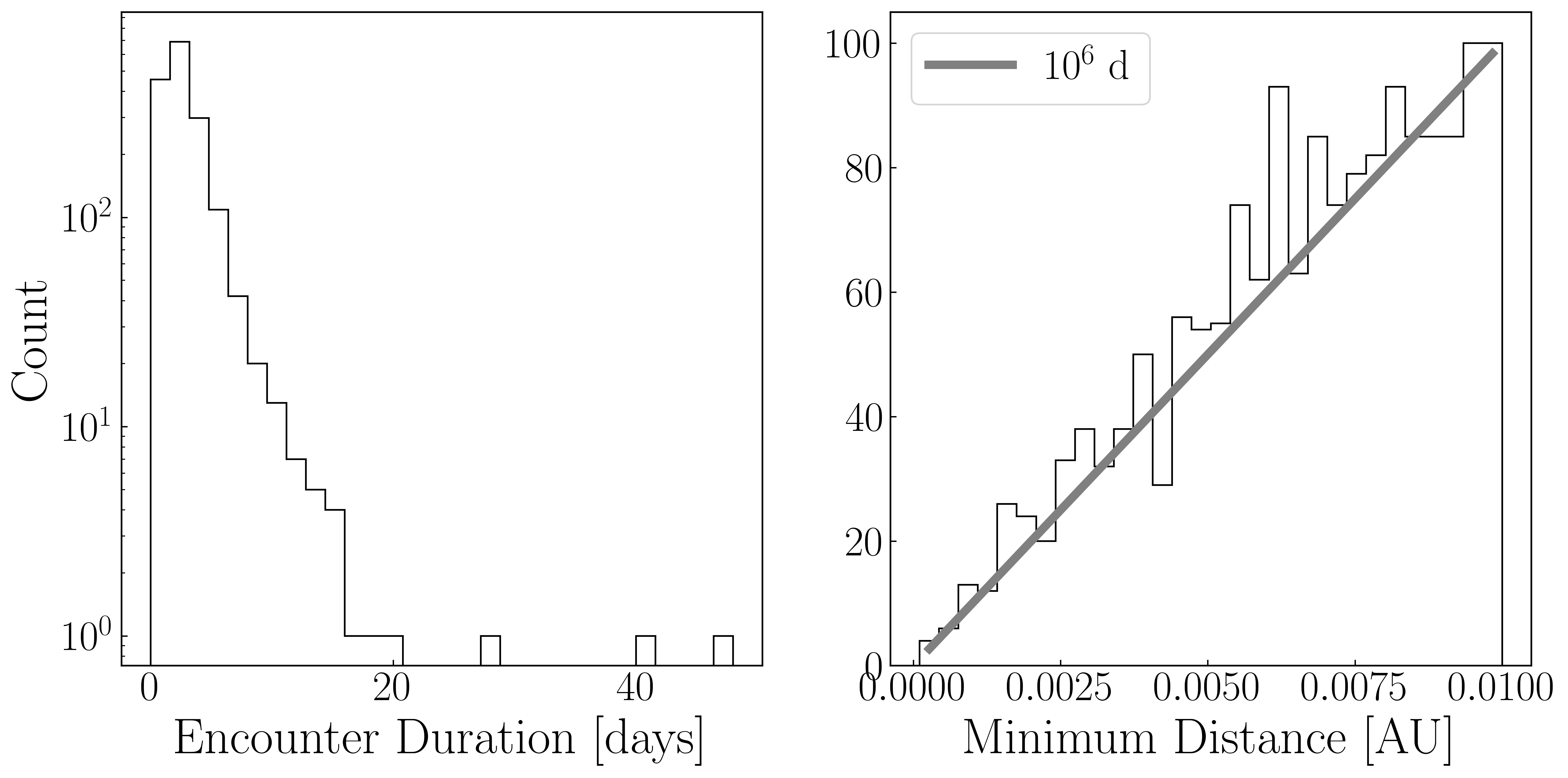}
    \caption{Here we show the distribution of parameters for our set of candidate encounters.  In the left panel, we show the distribution of encounter durations in days.  This distribution has a sharp peak near 5 days and a rapid fall-off for longer encounters. In the right panel, we show the distance of closest approach between the two bodies. There are linearly more grazing approaches than close approaches. This is consistent with what would be expected from a geometric cross-section.}
    \label{distributions}
\end{figure}

We can estimate the shape of the distribution of encounter distances using geometric arguments.  To approximate the distribution of minimum distances we may consider either asteroid to be at the center of a circular cross section.  An annulus on this cross section at distance $d$ from the asteroid will have area that is proportional to $d$.  Hence, there should be linearly more distant encounters than close encounters. This intuition is shown graphically in the summary of these encounters shown in Fig~\ref{distributions}.

%add trendline to plot.

\begin{table}
	\begin{tabular}{|lllrrrr|}
	\hline
	encounter & \textbf{mpc 1} & \textbf{mpc 2} & $t_0$ [JD] & $\delta$t [days] & d [$10^{-2}$ AU] & $\delta$v [$10^{-2}$ AU/day] \\
	\hline
	   1 & 54908 & 26001 & 2457001.2 & 0.9 & 0.848 & 1.085 \\
	   2 & 20452 & 26535 & 2457169.9 & 5.6 & 0.56 & 0.544 \\
      3 & 66223 & 11045 & 2457167.4 & 1.6 & 0.603 & 0.999 \\
	   4 & 1316 & 45387 & 2457082.8 & 1.5 & 0.571 & 1.046 \\
	   5 & 34708 & 6214 & 2457430.8 & 0.9 & 0.892 & 1.002 \\
	   6 & 3805 & 4513 & 2457410.4 & 3.2 & 0.392 & 0.758 \\
	   7 & 3840 & 16820 & 2457050.0 & 1.5 & 0.963 & 0.601 \\
	   8 & 17479 & 24920 & 2457321.5 & 2.6 & 0.086 & 0.875 \\
	   9 & 759 & 20730 & 2457111.4 & 1.1 & 0.624 & 1.192 \\
	   10 & 1058 & 4137 & 2457384.4 & 3.3 & 0.576 & 0.704 \\
	   11 & 4218 & 17170 & 2457352.2 & 2.9 & 0.694 & 0.704 \\
	   12 & 4912 & 19710 & 2457306.4 & 1.5 & 0.812 & 0.882 \\
	   13 & 4970 & 9780 & 2457039.6 & 5.8 & 0.329 & 0.571 \\
      14 & 1758 & 18815 & 2457499.7 & 1.0 & 0.967 & 0.711 \\	
	   15 & 38047 & 41424 & 2457231.8 & 0.9 & 0.471 & 1.400 \\
	\hline
	\end{tabular}
	
\caption{Here we show the essential parameter for our 15 strongest encounters. We show the encounter number (first column), the mpc numbers of the interacting objects (second and third column) the time of closest approach (fourth column), the encounter duration, and minimum distance (fifth and sixth columns) and the relative velocity between the bodies (final column). }
\label{table:encounters}
\end{table}

Having produced a catalog of potential encounters, we next determine which of these encounters produce observable astrometric deviations. The deflection from such an encounter is necessarily a function of the mass of the objects. However, since such masses are generally not known in advance, they must be estimated.  We produce our estimates in three steps. First, we check the JPL Small Bodies Database (SBDB) \footnote{\url{https://ssd.jpl.nasa.gov/tools/sbdb_lookup.html/}} to see if the asteroid already has a measured mass.  If it does, we assign the body that mass with a presumed 10\% bound on its error.  If no mass estimate is listed, we estimate one from the volume of the asteroid assuming a density in the range $0.5-4.5~\mathrm{g cm^{-3}}$.   This rather large range is chosen to encompass the possibility of high-density bodies like 16 Psyche as well as low-density rubble-pile asteroids.  To compute these volumes we assume spherical symmetric bodies with diameter given in the SBDB if available - again assuming 10\% error bounds.  If the diameter is not known, then it was estimated based on the asteroid's absolute magnitude and an assumed geometric albedo - with the albedo and error taken from \citet{Murray_2023}.  In the case that neither diameter nor albedo was available, the albedo was assumed to be 0.1 with 30\% error bounds.  This process effectively segments the asteroids into two main populations, one population which only is subject to uncertainty in density, and a second that is subject to both uncertainties in density and in diameter via the albedo.  

While equation \ref{eq:1} can be used to understand the strength of an encounter, it does not accurately predict the \textbf{physical} deviation of an asteroid from its unperturbed trajectory.  While the conversion from the deflection angle to physical deviation can be estimated in the limit of small eccentricities,  equation~\ref{eq:1} itself is only valid when the encounter velocity is nearly constant; this is not true for many of the longer encounters that appear in our sample. Hence, we determine the physical deviations due to the encounter by numerical integration.  We conduct three integrations for each encounter.  In the first, each body is given zero mass.  This integration serves as a control to which the others are compared.  In the second, both bodies are given the minimum mass in our estimated range.  In the third, each body is given the maximum estimated mass. By comparing these simulations we can numerically probe the range of physical deviations during the Gaia observing window.  Of these, only 15 have maximum potential deviations above $1 \mathrm{km}$ in any axis. These encounters are shown in Table~\ref{table:encounters}, which lists the essential parameters of the set of encounters. The encounter duration, JD of closest approach, and minimum distance are taken from our integrations, with the relative velocity estimated by $\delta v = 2 \sqrt{(0.01 \mathrm{AU})^2 + d^2} / \delta t$.  This small fraction suggests our initial cutoff of $0.01 \mathrm{AU}$ was large enough to capture all relevant encounters and that our set of 15 encounters is a complete list of strong scattering events that occur within the DR2 observing epoch. 

\begin{table}
	\begin{tabular}{|llllll|}
	\hline
	encounter & mpc & $x_{pet} - x_{unpet}$ [km] & $y_{pet} - y_{unpet}$ [km] & $z_{pet} - z_{unpet}$ [km] & $M$ [$10^{18}$ kg] \\
	\hline
	1 & 54908 & 0.0—0.0 & 0.01—0.1 & 0.02—0.23 & $4.2\cdot 10^{-3}$—$3.1\cdot 10^{-4}$ \\
	1 & 26001 & 0.76—8.81 & 0.01—0.06 & 0.01—0.09 & $3.\cdot 10^{-3}$—$2.6\cdot 10^{-4}$ \\
	\hline
	2 & 20452 & 0.0—0.0 & 0.0—0.0 & 0.0—0.0 & $4.2\cdot 10^{-4}$—$4.2\cdot 10^{-5}$ \\
	2 & 26535 & 0.29—3.13 & 0.2—2.2 & 0.22—2.4 & $3.5\cdot 10^{-3}$—$9.6\cdot 10^{-6}$ \\
	\hline
	3 & 66223 & 0.0—0.01 & 0.0—0.0 & 0.0—0.0 & $6.5\cdot 10^{-4}$—$2.0\cdot 10^{-5}$ \\
	3 & 11045 & 0.0—0.0 & 0.16—1.62 & 0.21—2.12 & $8.4\cdot 10^{-4}$—$4.\cdot 10^{-5}$ \\
	\hline
	4 & 1316 & 0.0—0.0 & 0.0—0.04 & 0.0—0.02 & $5.8\cdot 10^{-2}$—$4.3\cdot 10^{-5}$ \\
	4 & 45387 & 0.21—2.14 & 0.0—0.0 & 0.0—0.0 & $8.1\cdot 10^{-4}$—$6.2\cdot 10^{-5}$ \\
	\hline
	5 & 34708 & 0.0—0.55 & 0.0—0.13 & 0.0—1.54 & $5.6\cdot 10^{-4}$—$5.\cdot 10^{-5}$ \\
	5 & 6214 & 0.0—0.0 & 0.21—3.31 & 0.0—0.01 & $2.4\cdot 10^{-2}$—$2.3\cdot 10^{-3}$ \\
	\hline
	6 & 3805 & 0.0—0.0 & 0.0—0.0 & 0.0—0.0 & $6.8\cdot 10^{-3}$—$5.9\cdot 10^{-5}$ \\
	6 & 4513 & 0.0—0.0 & 0.01—0.08 & 0.19—2.88 & $1.3\cdot 10^{-2}$—$9.2\cdot 10^{-4}$ \\
	\hline
	7 & 3840 & 0.0—0.0 & 0.0—0.0 & 0.0—0.0 & $7.1\cdot 10^{-4}$—$6.6\cdot 10^{-5}$ \\
	7 & 16820 & 0.17—5.9 & 0.0—0.0 & 0.0—0.01 & $3.9\cdot 10^{-4}$—$2.8\cdot 10^{-5}$ \\
	\hline
	8 & 17479 & 0.0—0.0 & 0.0—0.0 & 0.0—0.0 & $2.1\cdot 10^{-3}$—$6.8\cdot 10^{-5}$ \\
	8 & 24920 & 0.16—1.68 & 0.0—0.03 & 0.0—0.0 & $5.4\cdot 10^{-4}$—$7.4\cdot 10^{-6}$ \\
	\hline
	9 & 20730 & 0.14—1.37 & 0.04—0.36 & 0.07—0.65 & $5.4\cdot 10^{-4}$—$5.3\cdot 10^{-5}$ \\
	9 & 759 & 0.0—0.0 & 0.0—0.0 & 0.0—0.0 & $7.2\cdot 10^{-1}$—$7.5\cdot 10^{-2}$ \\
	\hline
	10 & 1058 & 0.0—0.01 & 0.0—0.0 & 0.0—0.0 & $6.4\cdot 10^{-3}$—$6.5\cdot 10^{-4}$ \\
	10 & 4137 & 0.01—0.12 & 0.14—1.38 & 0.06—0.63 & $3.9\cdot 10^{-3}$—$2.6\cdot 10^{-5}$ \\
	\hline
	11 & 17170 & 0.02—0.22 & 0.12—1.27 & 0.06—0.61 & $2.2\cdot 10^{-3}$—$1.0\cdot 10^{-5}$ \\
	11 & 4218 & 0.0—0.0 & 0.0—0.0 & 0.0—0.0 & $1.3\cdot 10^{-3}$—$3.\cdot 10^{-6}$ \\
	\hline
	12 & 19710 & 0.01—0.29 & 0.09—1.85 & 0.05—1.08 & $1.8\cdot 10^{-3}$—$1.5\cdot 10^{-5}$ \\
	12 & 4912 & 0.0—0.0 & 0.0—0.0 & 0.0—0.0 & $1.4\cdot 10^{-3}$—$1.4\cdot 10^{-4}$ \\
	\hline
	13 & 9780 & 0.0—0.04 & 0.08—1.21 & 0.0—0.0 & $1.3\cdot 10^{-3}$—$7.2\cdot 10^{-5}$ \\
	13 & 4970 & 0.01—0.08 & 0.0—0.01 & 0.0—0.0 & $2.6\cdot 10^{-3}$—$2.6\cdot 10^{-4}$ \\
	\hline
	14 & 18815 & 0.01—1.21 & 0.0—0.43 & 0.0—0.01 & $8.2\cdot 10^{-3}$—$8.\cdot 10^{-4}$ \\
	14 & 1758 & 0.0—0.0 & 0.0—0.0 & 0.0—0.01 & $1.2\cdot 10^{-1}$—$1.2\cdot 10^{-3}$ \\
	\hline
	15 & 38047 & 0.0—0.0 & 0.0—0.0 & 0.0—0.0 & $8.4\cdot 10^{-5}$—$1.9\cdot 10^{-6}$ \\
	15 & 41424 & 0.0—0.0 & 0.01—1.72 & 0.0—0.03 & $1.5\cdot 10^{-4}$—$1.1\cdot 10^{-5}$ \\
	\hline
	\end{tabular}

\caption{ Predicted deviations in the position of the encountering bodies, expressed in km.  The upper limit comes from assuming the given asteroid has a large density and low albedo, the lower limit is derived making the opposite assumptions.  The upper and lower limits on the ranges of the approximate masses derived in this manner are shown in the rightmost column. In most cases, the majority of the scattering occurs on one of the bodies, but certain encounters scatter both objects comparably.  There is a great diversity in encounters with most scattering within the solar system plane. Scattering out of the plane, in the $z$ direction, is less common.}
\label{table:scattering}
\end{table}

\section{Modeling the Dynamics and Astrometry}
\label{sec:model}

Now that we have identified these 15 candidate encounters, we must model the dynamics of them with sufficient accuracy to estimate the masses, if possible. To do this we must ensure our dynamical model is at least as accurate as modeling Gaia's data requires.  Assuming the minimum possible measurement uncertainty, $1 \mathrm{mas}$, at the maximum possible distance from Gaia (about $3-4 \mathrm{AU}$), the angular uncertainty corresponds to a physical length scale of $\approx 1 \mathrm{km}$. Hence to ensure our simulations are not the limiting factor in our accuracy, we must ensure our integration of the asteroid trajectories is accurate $\mathcal{O} (1\mathrm{km})$ precision.  This requires we consider several different types of error, including integrator error, error in Gaia's state, errors due to light time corrections, errors due to photocenter displacement, and the limitations of the observation geometry.  In this section, we consider each effect.

The first priority for ensuring errors remain small over the duration of the integration is to verify that the physical model is sufficiently detailed to account for any effects that may produce deviations larger than $1 \mathrm{km}$ when applied over the 2 year Gaia DR2 time span.  To this end, we start with the model described earlier, consisting of the Sun, 8 major planets, and the 343 main-belt asteroids.  We then added general relativity using the \textit{gr} module in \texttt{REBOUNDx} package \citep{REBOUNDx}. The \textit{gr} module uses the post-Newtonian correction outlined in \citet{Newhall}, and assumes the dynamics are dominated by the force form a central body.  JPL uses a similar approach in the computation of the DE441 ephemeris \citet{Park_2021}. %equation 27
%we don't actually need the J2 to make htis work. 

\begin{figure}[!ht]
\centering
    \includegraphics[totalheight=10.5cm]{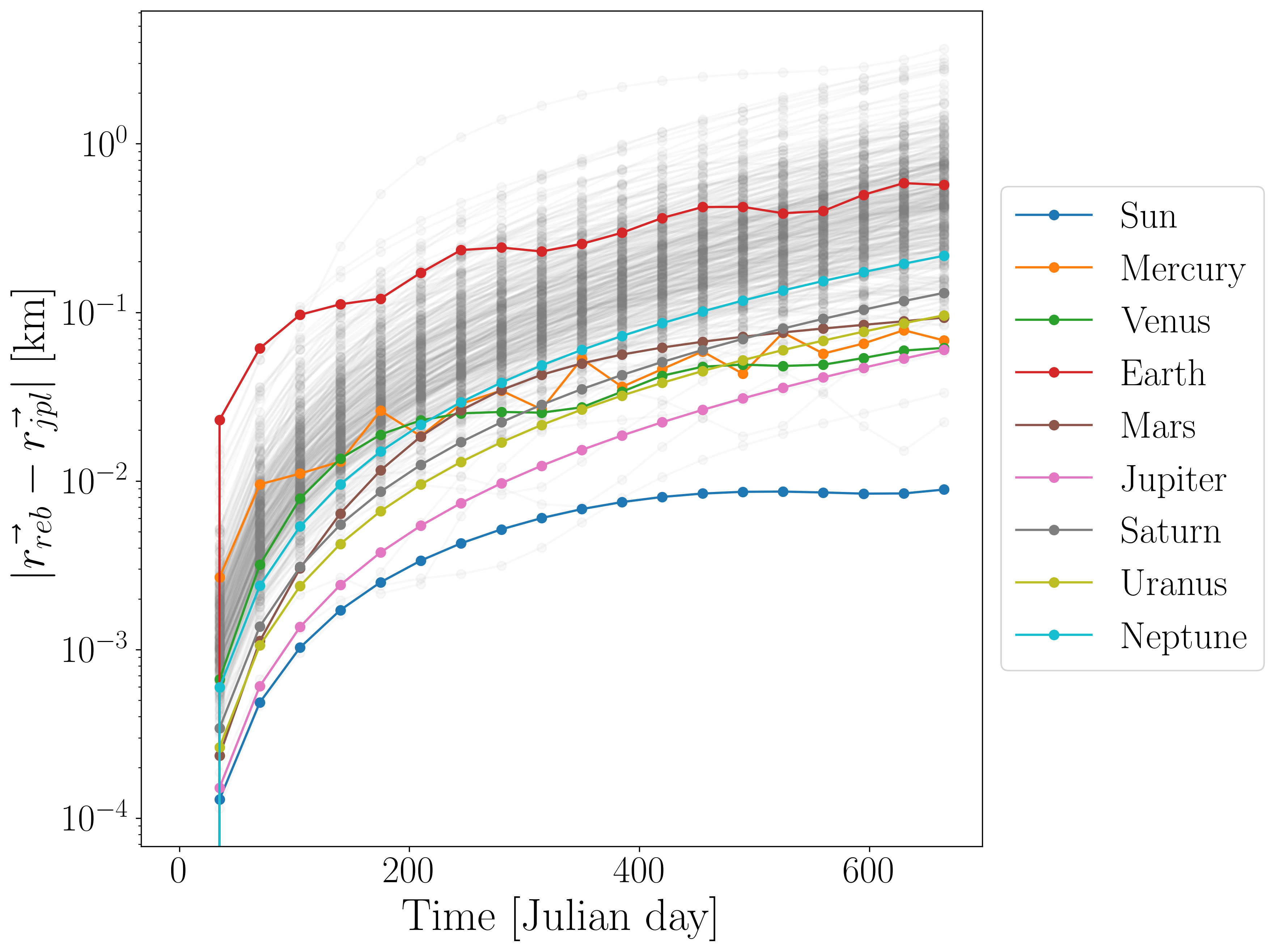}
    \caption{This figure shows the magnitude of the difference in heliocentric position between our integrations using \texttt{REBOUND} ($\vec{r}_{reb}$) and results from JPL Horizons ($\vec{r}_{jpl}$). The sun and major planets are labeled with specific colors. The 343 massive asteroids are shown in grey. We see that our integrator agrees to $\approx 1 \mathrm{km}$ level across the vast majority of bodies, exceeding our tolerance slightly for a few asteroids. The abnormally large error in the position of the earth is caused by differences in the modeling of the moon. }
    \label{agreement}
\end{figure}

To test the accuracy of our integrator, we compare our integrations to results from JPL Horizons.\footnote{\url{https://ssd.jpl.nasa.gov/horizons/app.html}} We begin these test integrations at JD 2456870.5 and continue them for $700~\mathrm{days}$. We use heliocentric rather than barycentric coordinates to eliminate any dependence on the specific model of the solar system between the two models. The difference between the heliocentric positions of these bodies produced by our integration to those recorded in JPL Horizons is shown in Fig \ref{agreement}. We find that over the time-span relevant for Gaia, our general relativistic integration agrees with JPL integrations to $ \approx 1 \mathrm{km}$ precision, but would start to cause a significant error if our integrations were to carry on several times longer.  We can therefore be confident that our integrations with this model do not artificially limit the precision with which we can determine scattering masses in our case, but integrator error may be a significant source of error for longer integrations needed in subsequent datareleases. 

% Insert Fig1. 

%on gaia's position error
To go from physical coordinates to astrometric predictions we must know the position of the Gaia spacecraft.  We obtain Gaia's position from the ESA Gaia Spice kernels, these positions are known to within about $30 \mathrm{m}$ in the radial direction and $300 \mathrm{m}$ along the other axes (De Albeniz, personal communication).  This uncertainty in Gaia's position is too small to affect our fits of scattering events for MBAs, but may be large enough to affect any scattering events observed in the future among closer bodies. 

% light-time displacement errors. 
In addition to position error, we must also take into account the effects of the finite speed of light on our astrometry. If $\vec{R}(t)$ is the position of the asteroid and $\vec{r}_{gaia}(t)$ is the position of Gaia, all at instant $t$.   We can then compute the light time delay through the recurrence relation 
\begin{equation}
    \tau_{n+1} = \frac{1}{c} \left[ \vec{R}(t-\tau_n)-\vec{r}_{gaia}(t) \right],
\end{equation}
where $c$ is the speed of light and $\tau_n$ is an approximation of the light time.  As $n \rightarrow \infty$, $\tau_n$ approaches the true light time.  We start with $\tau_0 = 0$, and within $5$ iterations the change in the apparent astrometry is on the order of $10^{-4}$ arcseconds, negligible compared to the other errors we consider.    
    
% Check potential photocenter displacement errors here.     
In addition to the light time correction, there is also an error that arises due to the photocenter displacement of the asteroids.  A body illuminated by the sun at some nonzero phase angle will not be uniformly illuminated. Therefore there will be some displacement between the observed astrometric position of the object and its center of mass. The photocenter displacement for a spherical body can be estimated at small phase angles by 
\begin{equation}
    d = \frac{3}{8} \phi R,
\end{equation}
where $d$ is the displacement, $\phi$ the phase angle, and $R$ the radius of the body \citep[e.g.][]{Aksnes_1986}.  This approximation is an excellent approximation to more complicated relationships derived from geometric or Lambertian scattering laws and is suitable for phase angles of $\approx 100^\circ$ \citep{Kaasalainen_2004}. For the 15 encounters in our filtered sample, the mean photocenter displacement for a fiducial phase angle of $20^{\circ}$ is 0.58 km.  This is smaller than the $\approx 1 \mathrm{km}$ limit on the Gaia astrometry. In addition, the most stringent bounds on masses will come from observations of the perturbed scattered body - not measurements of the primary.  These scattered bodies will be disproportionately small and will thus tend towards small photocenter displacements.  Hence, with our present level of accuracy, we ignore photocenter displacement when fitting our data. 

Finally, we must also account for the influence of the encounter geometry on the error budget. The state of each one of our asteroids comprises three position components and three velocity components.  However, each astrometric measurement has just two components, in the sky plane, and so individually cannot constrain the state. Additionally, as shown in Table (\ref{table:scattering}) the observable scattering can in general be oriented in arbitrary directions with respect to the coordinate axes, and so even strong deflections can be missed if they don't result in a angular displacement as seen from Gaia.  To make things worse, the Gaia astrometric process results in one of these directions being far less constrained than the other.  Consequently, each Gaia observation is effectively one-dimensional.  The combination of many of these observations may not be very constraining if the error ellipses are aligned with each other - but could work to constrain each other if oriented appropriately.  The orientation of the error ellipses on the sky is a complicated function of the time of observation, Gaia's position, and the location on the sky.  

In addition to orientation, the error ellipses have contributions from both random and systematic components.  The systematic component is constant for all positions in a given transit, the error is smallest for dim objects and increases for brighter objects.  This effect is due to several factors including CCD gating used to avoid saturation, and the increase in the apparent size of the asteroid introducing bias in the Gaia centroided positions. The random error has the opposite behavior, staying small for the brightest objects but increasing dramatically for objects dimmer than magnitude $m_G \approx 13.0$.  The total error is then a complicated function of the magnitude of the object being observed - itself a function of the time of observation and its orbital parameters. 

Due to the complicated nature of the error ellipse orientation and bounds, there's no good way to anticipate which of these encounters would be well constrained by Gaia data prior to fitting.  Since we have only $15$ potential encounters we instead simply fit every encounter and interpret the results.

\section{Fitting the Encounters}

%There is also potential to be errors due to un-modeled asteroids.  If there is an asteroid that is sufficiently massive it might perturb two other asteroids in a way to perturb their paths after a collision. 

    Once we have the astrometry we need to fit the scattering events. This task can be understood as a maximum likelihood problem and solved with sampling methods. However, integrating our relativistic model with all 343 massive asteroids is far too slow to run even with parallelized samplers.  Hence, if we wish to use sampling methods we must decide which objects matter most to our simulation.  We do this by running small simulations of each individual encounter using the same parameters as in Sec (\ref{sec:model}). In each of these simulations, we estimate the shift over the encounter $\delta r$ induced by each of our 343 massive asteroids on by numerically integrating their accelerations on the interacting bodies. We express the accelerations of the perturbing asteroid $i$ on interacting asteroid $j$ as  $a_{i,j} = \frac{G m_i}{(r_j - r_i)^2}$, and repeat this experiment for each encounter.  We take a conservative view and include the perturbing asteroid if, at any point in the simulation, $\delta r > 0.1 \mathrm{km}$.  This procedure always forces us to include the sun and 8 major planets, 1 Ceres, and $1-5$ other massive asteroids.  Integrating with just a few massive asteroids is much faster than integrating with the full 343 - this optimization results in simulations that are fast enough to be effectively used in a sampler.   In addition, to minimize our integrator errors even further, be begin our integrations in the middle of the Gaia observing epoch, and integrate both forward and backward.  While our integrator error will increase both forwards and backwards in time, each integration runs for only half as long which effectively halves the integrator error.

    In general, the likelihood function for a multivariable gaussian is 

    \begin{equation}
-        L_i = \det{2\pi K_i}^{1/2} \exp \left(-\frac{1}{2} (\vec{x_i}-\vec{\mu_i})^T K_i^{-1} (\vec{x_i}-\vec{\mu_i}) \right)
    \end{equation} 

    where $K_i$ is the covariance matrix of the observation and $\vec{\mu_i}$ specifies it's location on the sky. Here $i \in \{0,1\}$ can either specify the primary or secondary of the interaction.  The total log-likelihood, for a series of observations with errors of this form, can be written up to a constant term, as 
    \begin{equation}
    \log(L) = 
        \sum_{j}^{k_1} \left(-\frac{1}{2} (\vec{x_{1,j}}-\vec{\mu_{1,j}})^T K_{1,j}^{-1} (\vec{x_{1,j}}-\vec{\mu_{1,j}}) \right) + 
        \sum_{i}^{k_2} \left(-\frac{1}{2} (\vec{x_{2,j}}-\vec{\mu_{2,j}})^T K_{2,j}^{-1} (\vec{x_{2,j}}-\vec{\mu_{2,j}}) \right)
    \end{equation}
    
    where the sum is taken over the number of observations of each asteroid, (we assume $k_1$ observations of the primary and $k_2$ observations of the secondary). Expressing the likelihood in this way allows us to fit both encounters in which the primary is much larger than the secondary and encounters in which they have similar masses. This is important since many of our estimated masses in Sec (\ref{sec:model}) have large error bounds.  Here our covariance matrix $K$ is constructed from the sum of the right ascension and declination correlations and uncertainties, hence

\begin{gather}
 K
 =
 \begin{bmatrix} \sigma_{\alpha,r}^2 & \rho_r \sigma_{\delta,r} \sigma_{\alpha,r} \\ \rho_r \sigma_{\delta,r} \sigma_{\alpha,r} & \sigma_{\delta,r}^2 \end{bmatrix} + \begin{bmatrix} \sigma_{\alpha,s}^2 & \rho_s \sigma_{\delta,s} \sigma_{\alpha,s} \\ \rho_s \sigma_{\delta,s} \sigma_{\alpha,s} & \sigma_{\delta,s}^2 \end{bmatrix}.
\end{gather}

Where $\rho$ is the correlation between the errors in right ascension and declination.  The $r$ subscripted variables refer to the random components of the error and $s$ refer to the systematic component of the error.  Having defined our likelihood function, we can now find a range of model parameters that maximize it, subject to the astrometric uncertainties and prior knowledge about the asteroids. We fit these posteriors distributions using  \texttt{EMCEE} package described in \citet{Mackey_2013}. We fit a total of 14 variables; two are the masses of each of the bodies with the remaining 12 being shifts in the initial positions and velocities of the two bodies with respect to the JPL coordinates.  We adopt using uniform priors within $[\ -500, 500]\ \mathrm{km}$ in each of the positions, $ [\ -500 , 500 ]\ \mathrm{km/day}$ in velocity and $[\ 0 , 75 \cdot 10^{18} ]\ \mathrm{kg}$ in mass, these are chosen to be minimally informative, with the mass bounds an order of magnitude larger than the upper estimates for our largest bodies. This ensures our results are not strongly influenced by the prior and instead are a function of the Gaia data. 
The results of our fits are presented in Table (\ref{tab:results}).  The posteriors are approximately gaussian in both position and velocity and hence the $16^\mathrm{th}$ and $84^\mathrm{th}$ percentile bounds of the posteriors we quote can be understood as $1 \sigma$ bounds.  A few things are immediately clear upon examination of the results. First is that the initial conditions provided by JPL Horizons are generally excellent, with the vast majority of best-fit orbits lying within error tolerances. Second, velocities are incredibly well constrained, with the typical error of around $0.1 \mathrm{km/day}$.  Finally, we note that the combination of the geometric degradation in the measured astrometry and astrometric errors on constraints is significant.  Despite the best-case astrometry being accurate to $1 mas$, corresponding to spatial constraints of about a kilometer, few fits actually reach this tolerance. The sub-optimal Gaia astrometry, combined with incomplete phase space information and the differences in the geometry and epochs of observation result in real constraints that range from $1 \mathrm{km}$ to $20 \mathrm{km}$ scale accuracy and different substantially between different coordinate axes. 

\begin{figure}[!ht]
\centering
    \includegraphics[totalheight=9cm]{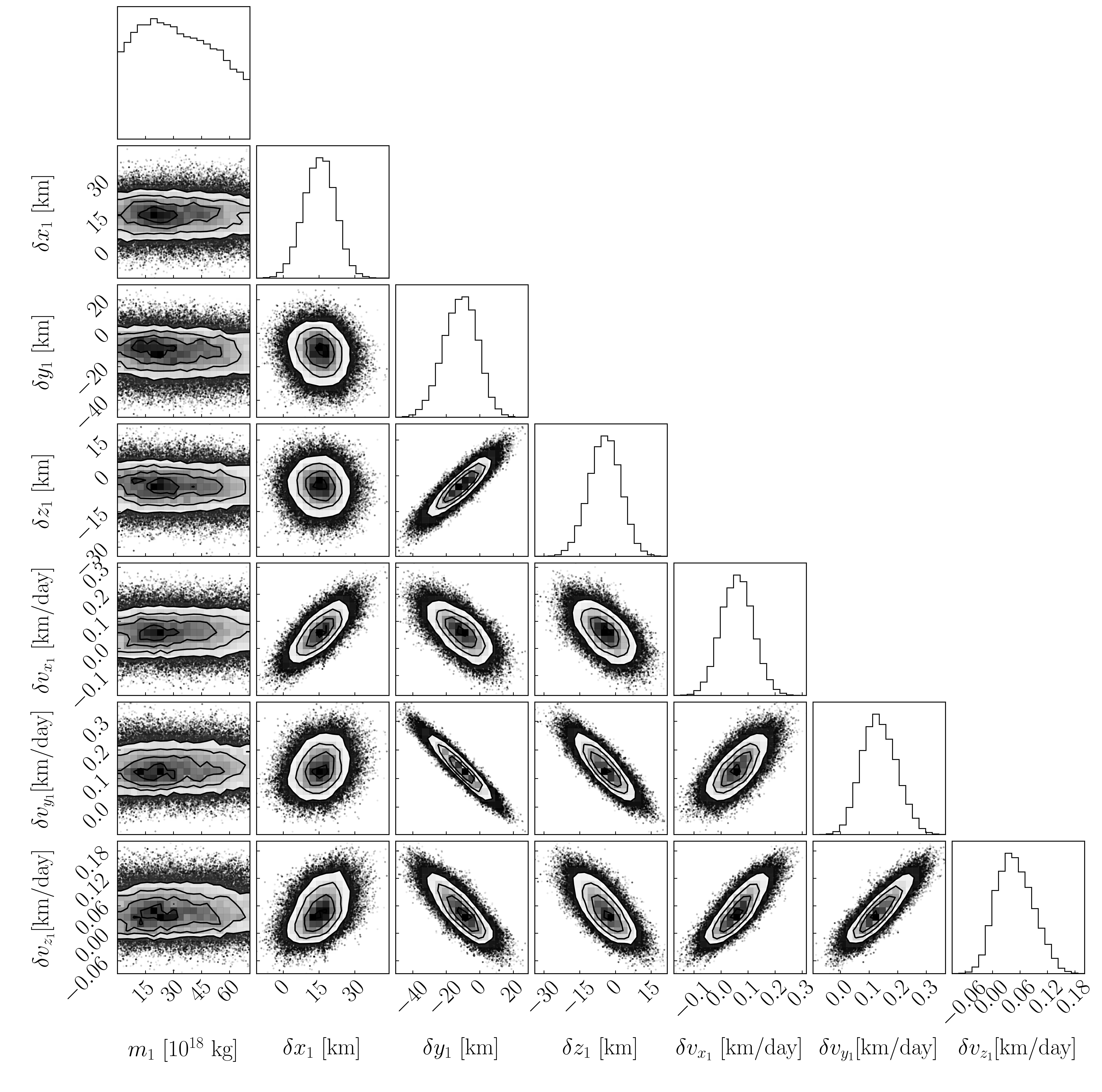}
    
    \includegraphics[totalheight=9cm]{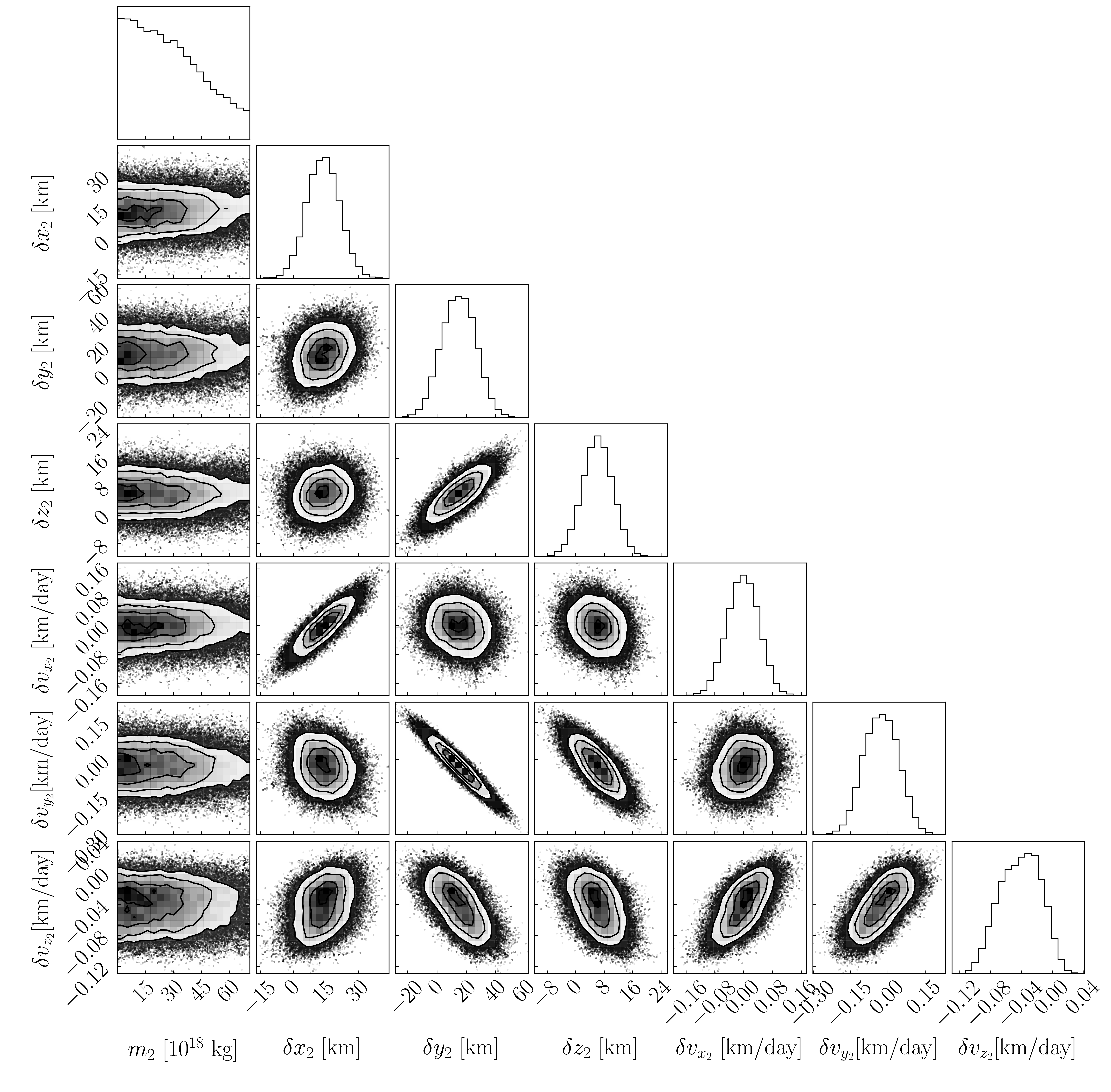}
    \caption{Here we show sections of a full corner plot for encounter 15, between asteroids 38047 and 41424. For clarity, we show only the correlations between the mass and state of each object. In this case, the state of one object is nearly uncorrelated with the mass of the other, hence little information is lost by not showing these entries. Evident are the significant correlations between the position and velocity of each asteroid. The $m_1$ posterior favors an unphysically large mass of $\approx 20 \cdot 10^{18} \mathrm{kg}$ whereas the $m_2$ posterior places a very loose upper bound on the mass of the secondary object.  Encounters like these are particularly ripe for follow-up with future observations}
    \label{fig:corner}
\end{figure}

\begin{figure}[!ht]
\centering
    \includegraphics[totalheight=20cm]{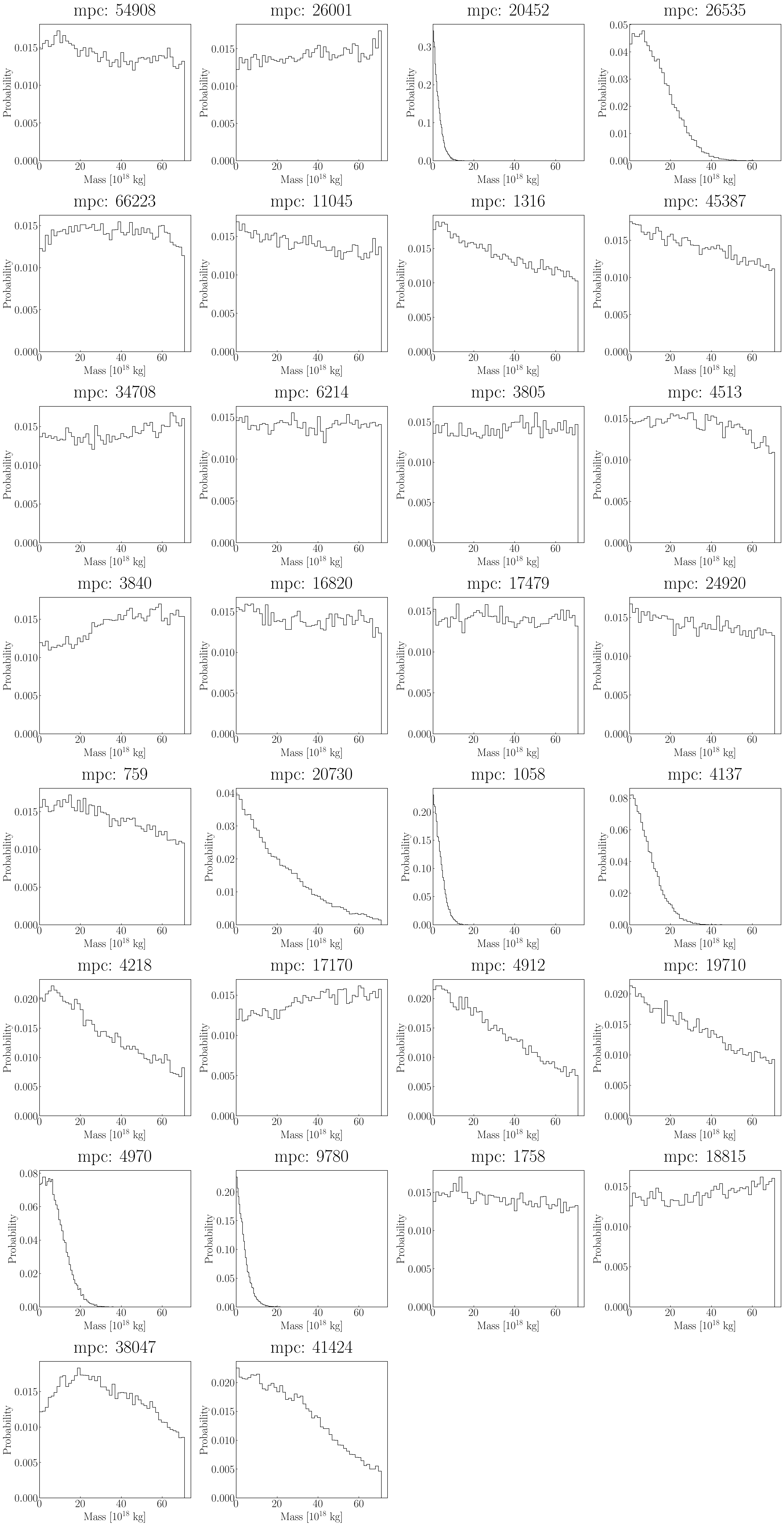}
    \caption{This figure shows a gallery of the posterior distributions of the masses derived for each encounter.  We can see in the majority of these encounters, no mass was significantly favored in our samples, and the mass is unconstrained. In a handful of cases (MPC 20452, 26535, 20730,1058,4137,4970 and 9780), large masses can be excluded to high confidence. There is only a single case, (MPC 38047) in which our finite, non-zero mass is preferred.  The cornerplot for this case is shown in Fig(\ref{fig:corner}).  }
    \label{fig:massposteriors}
\end{figure}

The posterior distributions in mass are a bit more complicated. Since the mass is constrained to be positive in the prior the posteriors resulting from our mass fit are highly non-gaussian, as can be seen in Fig(\ref{fig:corner}) and Fig(\ref{fig:massposteriors}), the percentile bounds cannot be understood as $1\sigma$ bounds.  In these cases, if the percentile bound were interpreted as the standard deviation of a normal distribution, the inferred distribution would be biased towards large masses - as there are no negative samples in the posterior.  This effect is a version of the well-known Lucy-Sweeny bias \citep{Lucy_1971}.  Since we have only a handful of strong encounters, we base our conclusions on direct visual examination of the posteriors.

\begin{table}
\centering
	\begin{tabular}{|lllllllll|}
	\hline
	encounter & mpc & $m$ [$10^{18}$ kg] & $\delta x$ [km] & $\delta y$ [km] & $\delta z$ [km] & $\delta v_{x}$ [km/day] & $\delta v_{y}$[km/day] & $\delta v_{z}$[km/day] \\
	\hline
	1 & 54908 & $35.20_{-24.48}^{+27.11}$ & $-5.95_{-4.52}^{+4.44}$ & $-7.00_{-21.24}^{+21.25}$ & $0.97_{-5.17}^{+5.18}$ & $0.01_{-0.05}^{+0.05}$ & $-0.44_{-0.49}^{+0.50}$ & $0.02_{-0.03}^{+0.03}$ \\
	1 & 26001 & $38.99_{-26.27}^{+24.64}$ & $3.39_{-4.59}^{+4.49}$ & $-6.94_{-20.32}^{+20.70}$ & $-8.15_{-4.92}^{+5.01}$ & $0.04_{-0.04}^{+0.04}$ & $-0.13_{-0.12}^{+0.12}$ & $-0.04_{-0.03}^{+0.03}$ \\
	\hline
	2 & 20452 & $1.91_{-1.40}^{+2.63}$ & $16.39_{-9.28}^{+9.74}$ & $-12.21_{-6.65}^{+6.39}$ & $-0.17_{-2.80}^{+2.90}$ & $0.07_{-0.05}^{+0.04}$ & $0.02_{-0.03}^{+0.03}$ & $-0.01_{-0.02}^{+0.02}$ \\
	2 & 26535 & $38.99_{-26.27}^{+24.64}$ & $3.39_{-4.59}^{+4.49}$ & $-6.94_{-20.32}^{+20.70}$ & $-8.15_{-4.92}^{+5.01}$ & $0.04_{-0.04}^{+0.04}$ & $-0.13_{-0.12}^{+0.12}$ & $-0.04_{-0.03}^{+0.03}$ \\
	\hline
	3 & 66223 & $37.22_{-24.75}^{+25.10}$ & $-12.86_{-10.82}^{+10.71}$ & $-1.65_{-2.90}^{+2.98}$ & $-12.36_{-7.24}^{+7.34}$ & $-0.02_{-0.03}^{+0.03}$ & $-0.03_{-0.04}^{+0.04}$ & $-0.03_{-0.03}^{+0.02}$ \\
	3 & 11045 & $38.99_{-26.27}^{+24.64}$ & $3.39_{-4.59}^{+4.49}$ & $-6.94_{-20.32}^{+20.70}$ & $-8.15_{-4.92}^{+5.01}$ & $0.04_{-0.04}^{+0.04}$ & $-0.13_{-0.12}^{+0.12}$ & $-0.04_{-0.03}^{+0.03}$ \\
	\hline
	4 & 1316 & $32.44_{-23.12}^{+27.62}$ & $-5.33_{-3.22}^{+3.24}$ & $35.29_{-24.82}^{+25.47}$ & $12.77_{-8.00}^{+8.15}$ & $-0.01_{-0.04}^{+0.04}$ & $-0.33_{-0.16}^{+0.17}$ & $-0.27_{-0.07}^{+0.07}$ \\
	4 & 45387 & $38.99_{-26.27}^{+24.64}$ & $3.39_{-4.59}^{+4.49}$ & $-6.94_{-20.32}^{+20.70}$ & $-8.15_{-4.92}^{+5.01}$ & $0.04_{-0.04}^{+0.04}$ & $-0.13_{-0.12}^{+0.12}$ & $-0.04_{-0.03}^{+0.03}$ \\
	\hline
	5 & 34708 & $39.14_{-26.76}^{+24.96}$ & $4.20_{-10.37}^{+9.90}$ & $16.95_{-5.03}^{+5.18}$ & $-0.33_{-3.89}^{+3.99}$ & $-0.06_{-0.05}^{+0.05}$ & $0.02_{-0.04}^{+0.03}$ & $-0.02_{-0.02}^{+0.02}$ \\
	5 & 6214 & $38.99_{-26.27}^{+24.64}$ & $3.39_{-4.59}^{+4.49}$ & $-6.94_{-20.32}^{+20.70}$ & $-8.15_{-4.92}^{+5.01}$ & $0.04_{-0.04}^{+0.04}$ & $-0.13_{-0.12}^{+0.12}$ & $-0.04_{-0.03}^{+0.03}$ \\
	\hline
	6 & 3805 & $38.41_{-26.32}^{+24.69}$ & $4.72_{-2.84}^{+2.77}$ & $20.30_{-6.40}^{+6.53}$ & $13.35_{-2.69}^{+2.76}$ & $0.00_{-0.02}^{+0.02}$ & $0.03_{-0.04}^{+0.04}$ & $0.03_{-0.02}^{+0.02}$ \\
	6 & 4513 & $38.99_{-26.27}^{+24.64}$ & $3.39_{-4.59}^{+4.49}$ & $-6.94_{-20.32}^{+20.70}$ & $-8.15_{-4.92}^{+5.01}$ & $0.04_{-0.04}^{+0.04}$ & $-0.13_{-0.12}^{+0.12}$ & $-0.04_{-0.03}^{+0.03}$ \\
	\hline
	7 & 3840 & $41.16_{-26.54}^{+22.88}$ & $-7.13_{-4.15}^{+4.28}$ & $9.16_{-6.39}^{+6.27}$ & $-1.95_{-6.42}^{+6.26}$ & $-0.04_{-0.03}^{+0.03}$ & $0.02_{-0.04}^{+0.04}$ & $-0.04_{-0.03}^{+0.03}$ \\
	7 & 16820 & $38.99_{-26.27}^{+24.64}$ & $3.39_{-4.59}^{+4.49}$ & $-6.94_{-20.32}^{+20.70}$ & $-8.15_{-4.92}^{+5.01}$ & $0.04_{-0.04}^{+0.04}$ & $-0.13_{-0.12}^{+0.12}$ & $-0.04_{-0.03}^{+0.03}$ \\
	\hline
	8 & 17479 & $36.94_{-24.83}^{+26.21}$ & $-17.73_{-5.95}^{+5.91}$ & $6.10_{-6.61}^{+6.62}$ & $-2.54_{-1.50}^{+1.56}$ & $0.03_{-0.05}^{+0.05}$ & $-0.05_{-0.05}^{+0.04}$ & $0.03_{-0.01}^{+0.01}$ \\
	8 & 24920 & $38.99_{-26.27}^{+24.64}$ & $3.39_{-4.59}^{+4.49}$ & $-6.94_{-20.32}^{+20.70}$ & $-8.15_{-4.92}^{+5.01}$ & $0.04_{-0.04}^{+0.04}$ & $-0.13_{-0.12}^{+0.12}$ & $-0.04_{-0.03}^{+0.03}$ \\
	\hline
	9 & 759 & $33.64_{-22.87}^{+26.54}$ & $9.27_{-4.73}^{+4.54}$ & $8.48_{-4.36}^{+4.54}$ & $-2.46_{-2.28}^{+2.28}$ & $0.03_{-0.02}^{+0.02}$ & $0.00_{-0.03}^{+0.03}$ & $-0.07_{-0.02}^{+0.02}$ \\
	9 & 20730 & $38.99_{-26.27}^{+24.64}$ & $3.39_{-4.59}^{+4.49}$ & $-6.94_{-20.32}^{+20.70}$ & $-8.15_{-4.92}^{+5.01}$ & $0.04_{-0.04}^{+0.04}$ & $-0.13_{-0.12}^{+0.12}$ & $-0.04_{-0.03}^{+0.03}$ \\
	\hline
	10 & 1058 & $2.69_{-1.94}^{+3.21}$ & $-1.19_{-8.29}^{+8.44}$ & $17.86_{-9.53}^{+8.99}$ & $1.59_{-3.11}^{+3.20}$ & $-0.03_{-0.06}^{+0.06}$ & $-0.01_{-0.05}^{+0.05}$ & $-0.05_{-0.01}^{+0.01}$ \\
	10 & 4137 & $38.99_{-26.27}^{+24.64}$ & $3.39_{-4.59}^{+4.49}$ & $-6.94_{-20.32}^{+20.70}$ & $-8.15_{-4.92}^{+5.01}$ & $0.04_{-0.04}^{+0.04}$ & $-0.13_{-0.12}^{+0.12}$ & $-0.04_{-0.03}^{+0.03}$ \\
	\hline
	11 & 4218 & $27.25_{-19.14}^{+28.40}$ & $-0.65_{-8.65}^{+8.78}$ & $-3.35_{-2.90}^{+2.87}$ & $6.49_{-2.58}^{+2.60}$ & $-0.15_{-0.12}^{+0.12}$ & $-0.02_{-0.05}^{+0.05}$ & $-0.03_{-0.04}^{+0.04}$ \\
	11 & 17170 & $38.99_{-26.27}^{+24.64}$ & $3.39_{-4.59}^{+4.49}$ & $-6.94_{-20.32}^{+20.70}$ & $-8.15_{-4.92}^{+5.01}$ & $0.04_{-0.04}^{+0.04}$ & $-0.13_{-0.12}^{+0.12}$ & $-0.04_{-0.03}^{+0.03}$ \\
	\hline
	12 & 4912 & $27.28_{-19.50}^{+27.35}$ & $-13.83_{-6.29}^{+6.67}$ & $32.22_{-10.04}^{+9.70}$ & $4.73_{-3.53}^{+3.51}$ & $-0.05_{-0.08}^{+0.08}$ & $-0.00_{-0.06}^{+0.06}$ & $-0.06_{-0.02}^{+0.02}$ \\
	12 & 19710 & $38.99_{-26.27}^{+24.64}$ & $3.39_{-4.59}^{+4.49}$ & $-6.94_{-20.32}^{+20.70}$ & $-8.15_{-4.92}^{+5.01}$ & $0.04_{-0.04}^{+0.04}$ & $-0.13_{-0.12}^{+0.12}$ & $-0.04_{-0.03}^{+0.03}$ \\
	\hline
	13 & 4970 & $7.03_{-4.76}^{+6.91}$ & $8.35_{-4.95}^{+5.07}$ & $3.79_{-4.43}^{+4.54}$ & $-10.32_{-2.87}^{+2.95}$ & $-0.02_{-0.02}^{+0.02}$ & $0.02_{-0.02}^{+0.02}$ & $-0.04_{-0.02}^{+0.02}$ \\
	13 & 9780 & $38.99_{-26.27}^{+24.64}$ & $3.39_{-4.59}^{+4.49}$ & $-6.94_{-20.32}^{+20.70}$ & $-8.15_{-4.92}^{+5.01}$ & $0.04_{-0.04}^{+0.04}$ & $-0.13_{-0.12}^{+0.12}$ & $-0.04_{-0.03}^{+0.03}$ \\
	\hline
	14 & 1758 & $35.99_{-24.56}^{+26.23}$ & $-9.50_{-7.77}^{+7.80}$ & $-2.56_{-2.15}^{+2.23}$ & $-7.94_{-2.23}^{+2.28}$ & $0.06_{-0.07}^{+0.07}$ & $-0.07_{-0.02}^{+0.02}$ & $0.00_{-0.03}^{+0.03}$ \\
	14 & 18815 & $38.99_{-26.27}^{+24.64}$ & $3.39_{-4.59}^{+4.49}$ & $-6.94_{-20.32}^{+20.70}$ & $-8.15_{-4.92}^{+5.01}$ & $0.04_{-0.04}^{+0.04}$ & $-0.13_{-0.12}^{+0.12}$ & $-0.04_{-0.03}^{+0.03}$ \\
	\hline
	15 & 38047 & $33.31_{-21.41}^{+25.16}$ & $15.35_{-6.73}^{+6.38}$ & $-11.27_{-10.39}^{+9.71}$ & $-4.41_{-6.69}^{+6.46}$ & $0.06_{-0.06}^{+0.06}$ & $0.13_{-0.06}^{+0.06}$ & $0.05_{-0.04}^{+0.04}$ \\
	15 & 41424 & $38.99_{-26.27}^{+24.64}$ & $3.39_{-4.59}^{+4.49}$ & $-6.94_{-20.32}^{+20.70}$ & $-8.15_{-4.92}^{+5.01}$ & $0.04_{-0.04}^{+0.04}$ & $-0.13_{-0.12}^{+0.12}$ & $-0.04_{-0.03}^{+0.03}$ \\
	\hline
	\end{tabular}
	
\caption{Here we summarize the results of our fits. We record, for each encounter the $50^\mathrm{th}$, $16^\mathrm{th}$ and $84^\mathrm{th}$ percentile bounds of the posterior distributions found by our sampler (an example of these distributions for a specific case can be found in Fig (\ref{fig:corner}). Each set of two rows corresponds to the fitted results from a single encounter.  The small shifts in the best-fit positions and velocities we derive is a testament to the accuracy of the JPL initial conditions and Gaia's precision. }
\label{tab:results}
\end{table}

The posteriors for the masses come in three main types. This first consists of distributions that are nearly uniform, such distributions impose no constraint on the masses of the interacting bodies these cases make up the majority of the cases in Fig(\ref{fig:massposteriors}).  In other cases, the posteriors are approximately half normal, with the peak of the probability mass function being near low masses.  In these examples, large masses are ruled out, but the masses of the interacting bodies are too small to be detected, the clearest examples include the posteriors resulting from fitting MPC 20452, 26535, 20730,1058,4137,4970, and 9780. In these cases, the $84^\mathrm{th}$ percentile of the distribution can be treated as an upper bound.  Finally, there is a single case where the dynamics impose a weak constraint on the mass, resulting in a peaked posterior distribution, but this peak is shifted away from the small masses predicted in Table(\ref{table:scattering}).  This is the case of MPC 38047, where uniquely, the posterior both deviates strongly from the uniform case and peaks at a large mass - several orders of magnitude too large to be plausibly attributed to the fitted body. An example of a corner plot for this case is shown in Fig (\ref{fig:corner}), where the posterior distribution of $m_1$, MPC 38047 can be compared to one more consistent with a smaller mass - that of $m_2$, MPC 41424.  Given this discrepancy, these deviations are unlikely to be due to a scattering process and may be the result of other unmodeled physics like non-gravitational forces.  While intriguing, the uncertainty in this case is still large, and so follow-up of this system with the data from Gaia's third data release will be needed to shed light on this encounter and to improve the constraint due to the increasing physical deviation in the trajectory. 

Due to the complicated uncertainties, including the astrometric precision of individual observations, the timing of those observations and the specific orbits of the bodies, it is difficult to estimate how much additional data would be necessary to constrain the masses of the objects we explore. However, in the typical case, where the uncertainties in position can be constrained to $\approx 5 \mathrm{km}$, doubling the number of observations, would increase the size of the perturbations roughly linearly, and shrink the uncertainty in position error by about $\sqrt{2}$.  Such observations will likely help constrain a large faction of the encounters we find in this work. However, we should still expect constraints of a variety of strength results depending on the specific orbits and observation time of the interacting bodies. In addition to future data releases, successor missions to Gaia, especially the LSST survey, also have the potential to further constrain the dynamics with astrometry, in addition to other methods that yield precise information about asteroid states, like occultations.  Finally, incorporating historical data may help constrain these encounters especially where systematic errors can be removed.

\section{Conclusion}

We find Gaia astrometry is an invaluable tool for studying scattering in the asteroid belt. However, significant obstacles remain to its use. Gaia's astrometry can be contaminated on the $1 mas$ level by several different sources, including photo-center displacement, uncertainty in the spacecraft's current position, and lensing corrections which will all have to be addressed to use the astrometry to its full potential for dynamical fitting.  In addition, the extreme precision of Gaia's astrometry requires a sophisticated dynamical model to not be artificially limited by integrator error.  A simple general relativistic implementation seems to be insufficient for time periods longer than $\approx 700 $ Julian Days, as after that time, the buildup of errors shown in Fig(\ref{agreement}) increase enough to be comparable with Gaia's precision.  Over timescales longer than this, more detailed integrations that incorporate, for example, the gravitational harmonics of the sun and planets may become necessary to fully exploit Gaia's precision. Finally, errors due to unmodeled massive asteroids or nongravitational effects may also be significant. 
Despite these challenges, we have shown that even a fairly basic model can place constraints on both the masses and dynamical state of asteroids in the main belt and that encounters even between relatively small asteroids can result in large enough deflections to be potentially detectable in the near future, making the objects we identify objects ripe for follow-up in future data releases. 

\section{Acknowledgements}

We are grateful to Matt Holman for helpful discussions and advice on minimizing integrator errors. We also thank the anonymous reviewers whose helpful advice greatly improved this paper.

\bibliography{refs}{}

\begin{thebibliography}{}
\expandafter\ifx\csname natexlab\endcsname\relax\def\natexlab#1{#1}\fi
\providecommand{\url}[1]{\href{#1}{#1}}
\providecommand{\dodoi}[1]{doi:~\href{http://doi.org/#1}{\nolinkurl{#1}}}
\providecommand{\doeprint}[1]{\href{http://ascl.net/#1}{\nolinkurl{http://ascl.net/#1}}}
\providecommand{\doarXiv}[1]{\href{https://arxiv.org/abs/#1}{\nolinkurl{https://arxiv.org/abs/#1}}}

\bibitem[{{Aksnes} {et~al.}(1986){Aksnes}, {Franklin}, \&
  {Magnusson}}]{Aksnes_1986}
{Aksnes}, K., {Franklin}, F., \& {Magnusson}, P. 1986, \aj, 92, 1436,
  \dodoi{10.1086/114280}

\bibitem[{{Baer} \& {Chesley}(2008)}]{Baer_2008}
{Baer}, J., \& {Chesley}, S.~R. 2008, Celestial Mechanics and Dynamical
  Astronomy, 100, 27, \dodoi{10.1007/s10569-007-9103-8}

\bibitem[{{Carry}(2012)}]{Carry_2012}
{Carry}, B. 2012, \planss, 73, 98, \dodoi{10.1016/j.pss.2012.03.009}

\bibitem[{{Danby}(1988)}]{Danby_1988}
{Danby}, J.~M.~A. 1988, {Fundamentals of celestial mechanics} (Scientific
  Research Publishing)

\bibitem[{{Everhart}(1985)}]{Everhart_1985}
{Everhart}, E. 1985, in Astrophysics and Space Science Library, Vol. 115, IAU
  Colloq. 83: Dynamics of Comets: Their Origin and Evolution, ed. A.~{Carusi}
  \& G.~B. {Valsecchi}, 185, \dodoi{10.1007/978-94-009-5400-7\_17}

\bibitem[{{Farnocchia} {et~al.}(2015){Farnocchia}, {Chesley}, {Chamberlin}, \&
  {Tholen}}]{Farnocchia_2015}
{Farnocchia}, D., {Chesley}, S.~R., {Chamberlin}, A.~B., \& {Tholen}, D.~J.
  2015, \icarus, 245, 94, \dodoi{10.1016/j.icarus.2014.07.033}

\bibitem[{{Folkner} {et~al.}(2014){Folkner}, {Williams}, {Boggs}, {Park}, \&
  {Kuchynka}}]{Folkner_2014}
{Folkner}, W.~M., {Williams}, J.~G., {Boggs}, D.~H., {Park}, R.~S., \&
  {Kuchynka}, P. 2014, Interplanetary Network Progress Report, 42-196, 1

\bibitem[{{Foreman-Mackey} {et~al.}(2013){Foreman-Mackey}, {Hogg}, {Lang}, \&
  {Goodman}}]{Mackey_2013}
{Foreman-Mackey}, D., {Hogg}, D.~W., {Lang}, D., \& {Goodman}, J. 2013, \pasp,
  125, 306, \dodoi{10.1086/670067}

\bibitem[{{Fortino} {et~al.}(2021){Fortino}, {Bernstein}, {Bernardinelli},
  {Aguena}, {Allam}, {Annis}, {Bacon}, {Bechtol}, {Bhargava}, {Brooks},
  {Burke}, {Carretero}, {Choi}, {Costanzi}, {Costa}, {Pereira}, {De Vicente},
  {Desai}, {Doel}, {Drlica-Wagner}, {Eckert}, {Eifler}, {Evrard}, {Ferrero},
  {Frieman}, {Garc{\'\i}a-Bellido}, {Gazta{\~n}aga}, {Gerdes}, {Gruendl},
  {Gschwend}, {Gutierrez}, {Hartley}, {Hinton}, {Hollowood}, {Honscheid},
  {James}, {Jarvis}, {Kent}, {Kuehn}, {Kuropatkin}, {Maia}, {Marshall},
  {Menanteau}, {Miquel}, {Morgan}, {Myles}, {Ogando}, {Palmese},
  {Paz-Chinch{\'o}n}, {Plazas}, {Roodman}, {Rykoff}, {Sanchez}, {Santiago},
  {Scarpine}, {Schubnell}, {Serrano}, {Sevilla-Noarbe}, {Smith}, {Suchyta},
  {Tarle}, {To}, {Tucker}, {Varga}, {Walker}, {Weller}, {Wester}, \& {DES
  Collaboration}}]{Fortino_2021}
{Fortino}, W.~F., {Bernstein}, G.~M., {Bernardinelli}, P.~H., {et~al.} 2021,
  \aj, 162, 106, \dodoi{10.3847/1538-3881/ac0722}

\bibitem[{{Fujiwara} {et~al.}(2006){Fujiwara}, {Kawaguchi}, {Yeomans}, {Abe},
  {Mukai}, {Okada}, {Saito}, {Yano}, {Yoshikawa}, {Scheeres}, {Barnouin-Jha},
  {Cheng}, {Demura}, {Gaskell}, {Hirata}, {Ikeda}, {Kominato}, {Miyamoto},
  {Nakamura}, {Nakamura}, {Sasaki}, \& {Uesugi}}]{Fujiwara_2006}
{Fujiwara}, A., {Kawaguchi}, J., {Yeomans}, D.~K., {et~al.} 2006, Science, 312,
  1330, \dodoi{10.1126/science.1125841}

\bibitem[{{Gaia Collaboration} {et~al.}(2016){Gaia Collaboration}, {Prusti},
  {de Bruijne}, {Brown}, {Vallenari}, {Babusiaux}, {Bailer-Jones}, {Bastian},
  {Biermann}, {Evans}, \& et~al.}]{GaiaCollaboration_2016}
{Gaia Collaboration}, {Prusti}, T., {de Bruijne}, J.~H.~J., {et~al.} 2016,
  \aap, 595, A1, \dodoi{10.1051/0004-6361/201629272}

\bibitem[{{Gaia Collaboration} {et~al.}(2018){Gaia Collaboration}, {Spoto},
  {Tanga}, {Mignard}, {Berthier}, {Carry}, {Cellino}, {Dell'Oro}, {Hestroffer},
  {Muinonen}, \& et~al.}]{GaiaCollaboration_2018}
{Gaia Collaboration}, {Spoto}, F., {Tanga}, P., {et~al.} 2018, \aap, 616, A13,
  \dodoi{10.1051/0004-6361/201832900}

\bibitem[{{Grundy} {et~al.}(2009){Grundy}, {Noll}, {Buie}, {Benecchi},
  {Stephens}, \& {Levison}}]{Grundy_2009}
{Grundy}, W.~M., {Noll}, K.~S., {Buie}, M.~W., {et~al.} 2009, \icarus, 200,
  627, \dodoi{10.1016/j.icarus.2008.12.008}

\bibitem[{{Hertz}(1966)}]{Hertz_1966}
{Hertz}, H.~G. 1966, \iaucirc, 1983, 3

\bibitem[{{Hilton}(2002)}]{Hilton_2002}
{Hilton}, J.~L. 2002, in Asteroids III (University of Arizona Press, Tucson),
  103--112

\bibitem[{{Hilton} {et~al.}(1996){Hilton}, {Seidelmann}, \&
  {Middour}}]{Hilton_1996}
{Hilton}, J.~L., {Seidelmann}, P.~K., \& {Middour}, J. 1996, \aj, 112, 2319,
  \dodoi{10.1086/118185}

\bibitem[{{Kaasalainen} \& {Tanga}(2004)}]{Kaasalainen_2004}
{Kaasalainen}, M., \& {Tanga}, P. 2004, \aap, 416, 367,
  \dodoi{10.1051/0004-6361:20031711}

\bibitem[{{Lucy} \& {Sweeney}(1971)}]{Lucy_1971}
{Lucy}, L.~B., \& {Sweeney}, M.~A. 1971, \aj, 76, 544, \dodoi{10.1086/111159}

\bibitem[{{Marchis} {et~al.}(2005){Marchis}, {Hestroffer}, {Descamps},
  {Berthier}, {Laver}, \& {de Pater}}]{Marchis_2005}
{Marchis}, F., {Hestroffer}, D., {Descamps}, P., {et~al.} 2005, \icarus, 178,
  450, \dodoi{10.1016/j.icarus.2005.05.003}

\bibitem[{{Murray}(2023)}]{Murray_2023}
{Murray}, Z. 2023, psj, 4, 90, \dodoi{10.3847/PSJ/acd381}

\bibitem[{{Newhall} {et~al.}(1983){Newhall}, {Standish}, \&
  {Williams}}]{Newhall}
{Newhall}, X.~X., {Standish}, E.~M., \& {Williams}, J.~G. 1983, \aap, 125, 150

\bibitem[{{Ostro} {et~al.}(2006){Ostro}, {Margot}, {Benner}, {Giorgini},
  {Scheeres}, {Fahnestock}, {Broschart}, {Bellerose}, {Nolan}, {Magri},
  {Pravec}, {Scheirich}, {Rose}, {Jurgens}, {De Jong}, \&
  {Suzuki}}]{Ostro_2006}
{Ostro}, S.~J., {Margot}, J.-L., {Benner}, L. A.~M., {et~al.} 2006, Science,
  314, 1276, \dodoi{10.1126/science.1133622}

\bibitem[{Park {et~al.}(2021)Park, Folkner, Williams, \& Boggs}]{Park_2021}
Park, R.~S., Folkner, W.~M., Williams, J.~G., \& Boggs, D.~H. 2021, The
  Astronomical Journal, 161, 105, \dodoi{10.3847/1538-3881/abd414}

\bibitem[{{Park} {et~al.}(2016){Park}, {Konopliv}, {Bills}, {Rambaux},
  {Castillo-Rogez}, {Raymond}, {Vaughan}, {Ermakov}, {Zuber}, {Fu}, {Toplis},
  {Russell}, {Nathues}, \& {Preusker}}]{Park_2016}
{Park}, R.~S., {Konopliv}, A.~S., {Bills}, B.~G., {et~al.} 2016, \nat, 537,
  515, \dodoi{10.1038/nature18955}

\bibitem[{{P{\"a}tzold} {et~al.}(2011){P{\"a}tzold}, {Andert}, {Asmar},
  {Anderson}, {Barriot}, {Bird}, {H{\"a}usler}, {Hahn}, {Tellmann}, {Sierks},
  {Lamy}, \& {Weiss}}]{Patzold_2011}
{P{\"a}tzold}, M., {Andert}, T.~P., {Asmar}, S.~W., {et~al.} 2011, Science,
  334, 491, \dodoi{10.1126/science.1209389}

\bibitem[{{Rein} \& {Spiegel}(2015)}]{Rein_2015}
{Rein}, H., \& {Spiegel}, D.~S. 2015, \mnras, 446, 1424,
  \dodoi{10.1093/mnras/stu2164}

\bibitem[{{Rein, H.} \& {Liu, S.-F.}(2012)}]{REBOUND}
{Rein, H.}, \& {Liu, S.-F.} 2012, A\&A, 537, A128,
  \dodoi{10.1051/0004-6361/201118085}

\bibitem[{{Russell} {et~al.}(2012){Russell}, {Raymond}, {Coradini}, {McSween},
  {Zuber}, {Nathues}, {De Sanctis}, {Jaumann}, {Konopliv}, {Preusker}, {Asmar},
  {Park}, {Gaskell}, {Keller}, {Mottola}, {Roatsch}, {Scully}, {Smith},
  {Tricarico}, {Toplis}, {Christensen}, {Feldman}, {Lawrence}, {McCoy},
  {Prettyman}, {Reedy}, {Sykes}, \& {Titus}}]{Russell_2012}
{Russell}, C.~T., {Raymond}, C.~A., {Coradini}, A., {et~al.} 2012, Science,
  336, 684, \dodoi{10.1126/science.1219381}

\bibitem[{{Shepard} {et~al.}(2006){Shepard}, {Margot}, {Magri}, {Nolan},
  {Schlieder}, {Estes}, {Bus}, {Volquardsen}, {Rivkin}, {Benner}, {Giorgini},
  {Ostro}, \& {Busch}}]{Shepard_2006}
{Shepard}, M.~K., {Margot}, J.-L., {Magri}, C., {et~al.} 2006, \icarus, 184,
  198, \dodoi{10.1016/j.icarus.2006.04.019}

\bibitem[{{Siltala} \& {Granvik}(2022)}]{Siltala_2022}
{Siltala}, L., \& {Granvik}, M. 2022, \aap, 658, A65,
  \dodoi{10.1051/0004-6361/202141459}

\bibitem[{{Stansberry} {et~al.}(2012){Stansberry}, {Grundy}, {Mueller},
  {Benecchi}, {Rieke}, {Noll}, {Buie}, {Levison}, {Porter}, \&
  {Roe}}]{Stansberry_2012}
{Stansberry}, J.~A., {Grundy}, W.~M., {Mueller}, M., {et~al.} 2012, \icarus,
  219, 676, \dodoi{10.1016/j.icarus.2012.03.029}

\bibitem[{Tamayo {et~al.}(2019)Tamayo, Rein, Shi, \& Hernandez}]{REBOUNDx}
Tamayo, D., Rein, H., Shi, P., \& Hernandez, D.~M. 2019, Monthly Notices of the
  Royal Astronomical Society, 491, 2885, \dodoi{10.1093/mnras/stz2870}

\bibitem[{{Yeomans} {et~al.}(1997){Yeomans}, {Barriot}, {Dunham}, {Farquhar},
  {Giorgini}, {Helfrich}, {Konopliv}, {McAdams}, {Miller}, {Owen}, {Scheeres},
  {Synnott}, \& {Williams}}]{Yeomans_1997}
{Yeomans}, D.~K., {Barriot}, J.~P., {Dunham}, D.~W., {et~al.} 1997, Science,
  278, 2106, \dodoi{10.1126/science.278.5346.2106}

\end{thebibliography}
\bibliographystyle{aasjournal}

\end{document}